\documentclass[12pt,preprint]{aastex}

\def \kms {$\,$km s$^{-1}$}

\def \arcsec {$^{\prime\prime}$}
\def \arcmin {$^{\prime}$}
\def \arcdeg {$^{\circ}$}

\def \kms {$\,$km s$^{-1}$}

\def \kms {km s$^{-1}$ }

\def \kms {$\,$km s$^{-1}$}

\long\def\symbolfootnote[#1]#2{\begingroup%
\def\thefootnote{\fnsymbol{footnote}}\footnote[#1]{#2}\endgroup}

\slugcomment{Accepted for Publication in The Astrophysical Journal}

\shorttitle{H$_2$CO Absorption toward Extragalactic Radio Sources} 
\shortauthors{Araya et al.}

\begin{document}

\title{Study of Interstellar Molecular Clouds using Formaldehyde Absorption toward Extragalactic Radio Sources}

\renewcommand{\thefootnote}{\fnsymbol{footnote}}

\author{E. D. Araya\altaffilmark{1},
N. Dieter-Conklin\altaffilmark{2},
W. M. Goss\altaffilmark{3},
N. Andreev\altaffilmark{1}}
\altaffiltext{1}{Physics Department, Western Illinois University, 
1 University Circle, Macomb, IL 61455, USA}
\altaffiltext{2}{Emeritus, Radio Astronomy Laboratory, 
University of California, Berkeley, USA}
\altaffiltext{3}{National Radio Astronomy Observatory, P.O. Box 0, Socorro, NM 87801, USA}

\renewcommand{\thefootnote}{\arabic{footnote}}\setcounter{footnote}{0}

\begin{abstract}

We present new Very Large Array 6$\,$cm H$_2$CO observations toward four
extragalactic radio continuum sources (B0212+735, 3C$\,$111, NRAO$\,$150, BL$\,$Lac) to 
explore the structure of foreground Galactic clouds as revealed by 
absorption variability. This project adds a new epoch in the monitoring observations of the
sources reported by Marscher and collaborators in the mid 1990's. Our new observations
confirm the monotonic increase in H$_2$CO absorption strength 
toward NRAO$\,$150. We do not detect 
significant variability of our 2009 spectra with respect to the 1994 spectra of 3C111, 
B0212+735 and BL$\,$Lac; however we find significant variability of the 3C111 2009 spectrum
with respect to archive observations conducted in 1991 and 1992.
Our analysis supports that changes in absorption lines could be caused by chemical 
and/or geometrical gradients in the foreground clouds, and not 
necessarily by small scale ($\sim 10\,$AU) high density molecular clumps within the clouds.
\end{abstract}


\section{Introduction}

The 6$\,$cm line of formaldehyde (H$_2$CO) is a powerful probe to 
study molecular gas in the Milky Way and in other 
galaxies (e.g., Mangum et al. 2008; Dickel et al. 2001). Soon after its
discovery (Snyder et al. 1969), the H$_2$CO 6$\,$cm transition
was detected in absorption toward radio continuum
sources as well as against the Cosmic Microwave Background 
(e.g., Palmer et al. 1969; Dieter 1973).
Now, more than 40 years of observations have shown that
the 6$\,$cm H$_2$CO line can trace large ($>5$pc) molecular 
clouds (e.g., Araya et al. 2007). Such large structures are not 
homogeneous, for example, Dieter-Conklin (2010) concluded that
molecular clouds traced by H$_2$CO absorption show different velocity 
dispersion when observed at effective resolutions greater or 
smaller than $\sim 0.5\,$pc, which gives direct evidence for sub-parsec 
structure traced by H$_2$CO.

Even smaller size-scales can be probed by H$_2$CO observations of 
absorption against compact extragalactic radio sources. 
In particular, Marscher et al. (1993) and Moore \& Marscher (1995)
used the VLA to monitor 6$\,$cm H$_2$CO absorption toward four extragalactic 
radio sources during a period of $3.4\,$years, and found evidence for
H$_2$CO line variability toward three continuum sources (3C$\,$111, NRAO$\,$150, 
BL Lac; they found no significant change in the B0212+735 line profile based on 
observations conducted less than 2 years apart). 
As recognized by Marscher et al. (1993) and Moore \& Marscher (1995),
the variability most likely traces changes in our line of sight 
that occurs because we move with respect to the 
distant source and intervening cloud, i.e., the combined effect of Earth's revolution 
around the Sun and the cloud's internal kinematics and proper motion (see 
Dieter-Conklin 2009, for a discussion of this ``searchlight effect'').
Marscher et al. (1993) estimated that the change in line of sight through the 
clouds enables probing physical scales of the order of $4\,$AU per year, 
i.e., $\sim 14\,$AU for sources with 3.4$\,$years monitoring data.

The use of 6$\,$cm H$_2$CO absorption as probe for substructure (as revealed by
spectral line variability) has clear advantages, in particular, the supercooling
effect of the lowest K-doublet transition due to H$_2$ - 
H$_2$CO collisions  (Townes \& Cheung 1969;
Thaddeus 1972; Garrison et al. 1975; Troscompt et al. 2009a) enhances the
optical depth against continuum sources. Also, the separation between the main group of five
hyperfine lines (centered near the $F = 2-2$ component) and the $F = 1-0$ component is 
approximately 18.2$\,$kHz (equivalent to 1.13\kms; see Tucker et al. 1970) 
and therefore the $F = 1-0$ line does not
significantly overlaps with the main hyperfine components in diffuse/translucent regions 
and dark clouds (which have linewidths of $\lesssim 1$\kms; see for example 
Araya et al. 2006). Thus, if variability is detected 
in the main absorption feature, variability of the hyperfine component is also expected, 
which is a useful test to assess whether artifacts are present in the data.

The monitoring of the H$_2$CO line profiles started by
Marscher and collaborators did not continue. In this paper we report new observations
of the sources observed by Moore \& Marscher (1995) to investigate longer term
variability in the absorption profiles and the implication of the variability
on the structure of the clouds.

\section{Observations}

We used the Very Large Array (VLA) of the 
National Radio Astronomy Observatory (NRAO) to observe the 
6$\,$cm line of H$_2$CO ($J_{KaKc} = 1_{10} - 1_{11}$, $\nu_0$ = 4829.6594$\,$MHz) 
in absorption toward the extragalactic radio sources
B0212+735 (J0217+738), 3C111 (J0418+380),
BL$\,$Lac (J2202+422), and NRAO150 (J0359+509). The observations were conducted
on 2009 March 2 (see Table~1)\footnote{VLA 
project number AA327.}.
The quasar J0319+415 was used as baseline/bandpass calibrator and 3C48
(J0137+331) as flux density calibrator (5.58$\,$Jy assumed at 4.8$\,$GHz). 
We measured a flux density of 13.8$\,$Jy for J0319+415.
We observed with 511 channels (channel width of 3.05$\,$kHz, 0.19\kms),
single circular polarization and a bandwidth of 1.56$\,$MHz
(96.8\kms), however, due to aliasing effects in the VLA-JVLA transition, 
only an effective bandwidth of 34\kms~was useful. The data reduction was
done in the NRAO software package AIPS, and the final spectra
were exported to CLASS\footnote{CLASS is part of the GILDAS software 
package developed by IRAM.} for analysis. 
Natural weighting was used to create the images to maximize
signal-to-noise. All sources were self-calibrated. 
Details of the observation setup and
synthesized beam are given in Table~1.

In the case of 3C111, we conducted a short (10 min) observation in 
standard VLA C-Band continuum mode (4IF; 50 MHz per IF). 
J0414+343 was used as complex gain calibrator. We measured a flux 
density of 1.4$\,$Jy for J0414+343. The radio continuum image of 
3C111 was used as initial model for the self-calibration of the 
3C111 line data.

We downloaded from the VLA archive the raw-data of 
previous H$_2$CO observations of all our sources, which
have been already published by Marscher et al. (1993) and 
Moore \& Marscher (1995). We reduced the archive data in AIPS following 
standard procedures. The spectra were Hanning smoothed to a final 
channel width of 3.05$\,$kHz (0.19\kms). Table~1 also lists the details 
of the archive data discussed in this paper.

\section{Results}

We detected radio continuum and H$_2$CO absorption toward the four 
sources. Table~2 lists the line parameters of the 
H$_2$CO detections obtained
from single-Gaussian fits. No significant variability was detected
toward B0212+735 and BL$\,$Lac (Figures~1 and 2, respectively).
In the upper left panels of Figures~1 and 2, we show the radio continuum 
images (contours) and the peak-pixel absorption images (grey scale). 
In the upper right panels of Figures~1 and 2, we show
the March 2009 spectra, including a fit to the six hyperfine components 
(blue dashed lines; Table~3). The lower left sequence of panels 
show the March 2009 spectrum (in black) and the spectra from the VLA archive 
obtained at previous epochs (in red). The lower right sequence of panels
show the difference between the previous observations and the 2009
data. In contrast to B0212+735 and BL$\,$Lac, significant variability was detected 
toward NRAO$\,$150 (Figure~3).

These three sources (B0212+735, BL$\,$Lac, NRAO$\,$150) show compact 
radio emission. In contrast, 3C111 was resolved in a central compact source (Core), and 
two extended radio lobes to the NE and SW
(see Figure~4). The H$_2$CO spectra and hyperfine fits of the NE and
SW components are also shown in Figure~4. The low signal-to-noise
toward the NE and SW regions precludes meaningful analysis of
time variability, thus, only the 2009 spectra
are shown. The signal-to-noise toward the 3C111 Core is as good as 
that of B0212+735, NRAO$\,$150 and BL$\,$Lac, and variability is detected
(Figure~5).

\section{Analysis and Discussion}

\subsection{Reproducibility of Spectral Line Results}

To check the effect of different calibration procedures, we reduced all 
archive observations of the sources which have been
already published by Marscher et al. (1993) and Moore \& Marscher (1995). 
We applied completely independent flagging, calibration and imaging procedures. 
In our analysis of the physical parameters of the clouds, the peak optical 
depths and linewidths are input parameters (see section 4.2), while Moore \& Marscher (1995) 
used equivalent widths defined as $\int \tau dv$. To compare our results
to those of Moore \& Marscher (1995), we measured the equivalent widths of our 
reduction of the M94 data. Our equivalent widths agree with Moore \& Marscher (1995) values 
within statistical uncertainties from the fit in all sources, which shows that differences
in calibration procedures are not a significant source of error in determining 
variability. For example, in the case of the M94 observations of B0212+735, 
Moore \& Marscher (1995) reported $\int \tau dv$ = 0.0893, $V_{LSR, F=2-2} = 3.56$\kms, 
and $FWHM_{F=2-2} = 0.84$\kms, while we measured $\int \tau dv = 0.091 \pm 0.012$, 
$V_{LSR, F=2-2} = 3.57 \pm 0.01$\kms, and $FWHM_{F=2-2} = 0.81 \pm 0.04$\kms.

We detect a strong absorption line at $-17.2$\kms~toward NRAO$\,$150 
(see Table~3, Figure~3). This line 
was included in the bandpass (different IFs) of the J91 and D90
epochs, however, the line was not included in the bandpass of the M94 and D92 epochs. 
As clearly seen in Figure~3, the line
profile of the $-17.2$\kms~component is perfectly well reproduced 
by our 2009 observations with respect to the previous epochs, 
showing that the variability of the $-10.5$\kms~component in NRAO$\,$150 is not an artifact.
Likewise, the $-2.3$\kms~velocity component in 3C111 shows no variability 
(compare red and black spectra in Figure~5) while the main absorption feature 
at $-1.0$\kms~does. 

Careful inspection of the B0212+735 and NRAO$\,$150 spectra (Figures~1 and 3) shows 
that the bandpasses are not perfectly flat; specifically, the B0212+735 spectrum has 
a weak absorption feature at $\sim$0\kms~and NRAO$\,$150 presents weak absorption features
at $-$14 and $-$4.5\kms~(highlighted with arrows). 
Absorption features at these velocities have been seen before in
other molecular transitions. The CO, OH, and HCO$^+$ spectra 
of B0212+735 reported by 
Liszt \& Pety (2012) show a velocity component at $\sim 0$\kms~in 
addition to the 
main component at $\sim 3.5$\kms. In the case of NRAO$\,$150, 
molecular absorption
features at $-$14 and $-$4.5\kms~(in addition to the components at $-17.3$ and
$-10.5$ and $-8.5$\kms) have been reported by Pety et al. (2008) in CO, HCO$^+$, C$_2$H,
and HNC; of all, the HNC spectrum is the most similar to the H$_2$CO spectrum.
The same weak absorption features are seen at the other epochs (see Figure~3, 
left spectra). The remarkable reproducibility of these weak features demonstrates the 
superb stability of the VLA for long-term variability studies of spectral lines.

\subsection{Physical Parameters}

We used the {\tt Molpop-CEP} code (Elitzur \& Asensio Ramos 2006) 
to explore the physical conditions of the foreground clouds 
based on the observed H$_2$CO absorption lines. 
We used the H$_2$CO -- para-H$_2$ collision rates of
Troscompt et al. (2009b), and included He collisions using the 
rates by Green (1991)\footnote{Electron collisions could be an
important factor in the H$_2$CO excitation in translucent/diffuse clouds
(e.g., Turner 1993), however development of 
a model including electron collisions
is beyond the scope of this work.}. Diffuse/translucent clouds 
have densities between
$\sim 10^2$ and $\sim 10^4\,$cm$^{-3}$ (e.g., van Dishoeck \& Black 1988), 
and according to Moore \& Marscher (1995) may contain even higher 
density clumps [$\sim 10^6\,$cm$^{-3}$; however see Liszt \& Lucas (2000) and our
discussion below]. The H$_2$CO abundance relative to hydrogen may be 
somewhere between $10^{-9}$ and $\sim 5 \times 10^{-8}$; 
for example, Zhou et al. (1990) estimated an 
abundance of $2 \times 10^{-9}$ toward B335; Turner (1993) report an 
abundance of $6 \times 10^{-9}$; 
Troscompt et al. (2009a) obtained an abundance of $6 \times 10^{-8}$.
We ran grids of models at two kinetic temperatures (10 and 30$\,$K), 
in a range of densities between 10$^2$ to 10$^6\,$cm$^{-3}$ and H$_2$CO 
abundance ratios between $10^{-9}$ and $5 \times 10^{-8}$, 
however, for the analysis, we further constrain the parameters as follows:

{\it Temperature:} The extinction toward the 
extragalactic sources discussed here is $A_V \sim 1\,$mag or greater  
(e.g., Marscher et al. 1993; Schlegel et al. 1998; Pety et al. 2008; 
Schlafly \& Finkbeiner 2011; Liszt \& Pety 2012). Hence, the foreground 
clouds are in the diffuse/translucent boundary; which 
are expected to be hotter than $\sim$10$\,$K at their centers and as hot as 
$\sim 50\,$K at their boundaries (e.g., van Dishoeck \& Black 1988). 
Given that the H$_2$CO molecules are not expected at the boundaries of the clouds
but at higher depths where they may be shielded from the interstellar 
radiation field, a temperature of $\sim 30\,$K is assumed.  

{\it Density:} observations of 6$\,$cm H$_2$CO absorption profiles in a number of
environments including dark nebulae, pre-protostellar cores and 
low mass star forming regions with absorption against the CMB (e.g., 
Palmer et al. 1969; Heiles 1973; Dieter 1973; Young et al. 2004; 
Araya et al. 2006) have resulted in excitation temperature measurements
of the 6$\,$cm K-doublet of $\lesssim 2\,$K, which according to our model
imply densities below $\sim 10^5\,$cm$^{-3}$. This is the well known
result that the 6$\,$cm H$_2$CO K-doublet thermalizes at densities of
$\sim 10^6\,$cm$^{-3}$ (e.g., Troscompt et al. 2009a; Mundy et al. 1987).
Thus, most of the H$_2$CO gas likely traces regions of densities below
$\sim 10^5\,$cm$^{-3}$.
A lower limit of the density can be inferred from our data as well:
as highlighted in Figures~1 and 3 upper-right panels and discussed
above, we detected some weak H$_2$CO absorption features. These velocity 
components are quite prominent in CO and HCO$^+$ (e.g., Pety et al. 2008), and may
trace molecular gas at densities of $\sim 10^2\,$cm$^{-3}$ (e.g., 
Liszt \& Lucas 2000; Pety et al. 2008). H$_2$CO is likely weaker because
higher density conditions would be needed for stronger H$_2$CO absorption 
features (Pety et al. 2008). Thus, the H$_2$CO gas likely traces
material at densities between $\sim 10^3$ to $10^5\,$cm$^{-3}$, 
i.e., higher than those traced by HCO$^+$ and lower than termalization
densities; we therefore assume a density of $10^4\,$cm$^{-3}$ for the 
analysis.

{\it Abundance:} we simply assume an abundance of $5\times10^{-9}$,
i.e., similar to the abundance reported by Turner (1993) for high-latitude cirrus clouds.

Based on these assumptions, we can calculate the H$_2$CO and H$_2$
column densities, and the thickness ($\Delta L$) of the absorbing region 
(assuming a slab geometry), i.e., the depth of the cloud that would result in the
observed absorption spectrum for a given density, temperature and 
abundance. We note that the slab thickness should
be considered an effective thickness, more precisely, a lower limit 
of the physical path which would differ from the true thickness of 
the cloud depending on the number of molecular clumps and their 
separation along the line of sight (see Marscher et al. 1993).

In the case of the H$_2$CO absorption region toward the core of 3C111 (see Figure~4), 
we obtained the following parameters and uncertainty ranges:
$\Delta L = 1.0 \times 10^4\,$AU ([$7 \times 10^2$, $5 \times 10^5$] AU),
$N(H_2) = 1.5 \times 10^{21}\,$cm$^{-2}$ ([$1 \times 10^{20}$, $5 \times 10^{22}$] cm$^{-2}$), and
$N(H_2CO) = 7.6 \times 10^{12}\,$cm$^{-2}$ ([$7 \times 10^{12}$, $5 \times 10^{13}$] cm$^{-2}$), 
where the ranges shown in square brackets were obtained by 
assuming densities between $10^3$ and $10^5\,$cm$^{-3}$ and formaldehyde
abundances with respect to hydrogen between $10^{-9}$ and $5 \times 10^{-8}$.
We note that H$_2$CO column densities of $\sim 7\times 10^{12}\,$cm$^{-2}$ have been
reported in diffuse clouds, including the ones discussed here (see 
Liszt et al. 2006). 

Even though the value of $\Delta L$ is not well constrained within
a density range of $10^3$ to $10^5\,$cm$^{-3}$, we can assess how
reasonable the $\Delta L = 1.0 \times 10^4\,$AU value is.
As pointed out by Moore \& Marscher (1995), the extended brightness
distribution of 3C111 enables a study of spatial changes 
of the H$_2$CO absorption by comparing spectral lines toward the 
NE, Core, and SW continuum sources. Detection of absorption along the
three sight-lines shows that the overall cloud is more extended than the
projected separation of the NE and SW components, i.e., 
$7\times 10^4\,$AU (0.3$\,$pc) assuming a distance of 350$\,$pc (Moore \& 
Marscher 1995; obtained from a photometric/extinction analysis by 
Ungerechts \& Thaddeus 1987)\footnote{For consistency with Moore \& Marscher (1995) we use
the same distance they assumed for the foreground cloud, however, the possible
association with the California Molecular Cloud would imply a distance of 
$\sim$450$\,$pc (Harvey et al. 2013).}. A $\Delta L = 1.0 \times 10^4\,$AU 
would imply that the depth of the cloud is somewhat less than the 
extent of the cloud in the plane of the sky, which is consistent with having 
slightly different line parameters toward the different continuum 
sources (Table~3)\footnote{We note that velocity components at $-$1\kms~and
$\-2$\kms~were detected toward the Core and the NE regions,
whereas only a single broad line was detected in our lower signal-to-noise 
spectrum of the SW 3C111 region. Thus, the significantly different 
H$_2$CO SW 3C111 profile could be caused in part due to blended lines in the
low signal-to-noise spectrum.}.

Assuming the same values of kinetic temperature (30$\,$K), 
density ($10^4\,$cm$^{-3}$), H$_2$CO abundance ($5\times10^{-9}$), and abundance/density 
domains as above, we obtain a total molecular column density 
and slab thickness equal to $N(H_2) = 3 \times 10^{20}\,$cm$^{-2}$ 
([$3 \times 10^{19}$, $5 \times 10^{21}$] cm$^{-2}$)
and $\Delta L = 2 \times 10^3\,$AU ([$6 \times 10^1$, $1 \times 10^{5}$] AU) for B0212+735. 
In the case of NRAO$\,$150 we obtain 
$N(H_2) = 6 \times 10^{20}\,$cm$^{-2}$ ([$6 \times 10^{19}$, $1 \times 10^{22}$] cm$^{-2}$)
and $\Delta L = 4 \times 10^3\,$AU ([$2 \times 10^2$, $2 \times 10^{5}$] AU), 
while for BL$\,$Lac we obtain 
$N(H_2) = 2 \times 10^{20}\,$cm$^{-2}$ ([$2 \times 10^{19}$, $2 \times 10^{21}$] cm$^{-2}$)
and $\Delta L = 10^3\,$AU ([$3 \times 10^1$, $5 \times 10^{4}$] AU).

\newpage

\subsection{Physical Scales Probed by the Monitoring Observations}

The most important aspect to relate absorption variability to cloud 
structure is to be able to convert 
time-lapses to physical displacement of the clouds perpendicular to our 
sight line, i.e., the proper motion due to the combined effect of Earth's 
revolution and the motion of the Solar System and the cloud around the Galaxy. 
Marscher et al. (1993) and Moore \& Marscher (1995) 
estimated that the magnitude of transverse displacement for the 
foreground clouds toward NRAO$\,$150, 3C111 and BL Lac is of 
the order of $\sim 4\,$AU per year. Specifically, 
Moore \& Marscher (1995) report that after approximately 3.4 years of
monitoring observations, the foreground clouds toward NRAO$\,$150
and BL Lac would have had transverse displacements of 13.9$\,$AU and 
12.4$\,$AU, respectively.
These estimates depend on the precise knowledge 
of the transverse velocity
of the foreground clouds with respect to Earth, which has not been 
directly measured. Thus, we recalculated the transverse
displacement of the foreground clouds including uncertainty in
the cloud's peculiar velocity around the Galaxy, in addition to the
line-of-sight displacements due to Earth's revolution, LSR reference
frame correction, peculiar velocity of the Solar System and estimated
Galactic circular rotation of the foreground clouds. To calculate the 
Galactic circular rotation, we assumed the distances reported by 
Moore \& Marscher (1995), i.e., 700$\,$pc 
(kinematic distance, Marscher et al. 1993), 350$\,$pc (estimated 
from a photometric/extinction analysis by Ungerechts \& Thaddeus 1987), 
and 330$\,$pc (estimated from the average vertical height of Galactic 
molecular clouds in the solar neighborhood and the galactic latitude of the 
foreground cloud, Lucas \& Liszt 1993)
for the NRAO$\,$150, 3C111, and BL$\,$Lac foreground clouds, respectively. 
We used the rotation curve of Clemens (1985) for consistency with 
Marscher et al. (1993). To estimate the range of possible transverse
displacements, we assume a peculiar velocity of the foreground clouds
of $\pm 10$\kms~in the three galactocentric cylindrical coordinates. 
In a time lapse of 748 days [the time lapse of the Marscher et al. (1993) 
observations], we obtain a maximum displacement of $\sim$10$\,$AU 
and a minimum of $\sim 0\,$AU depending on the peculiar velocity of the clouds 
toward NRAO$\,$150, 3C111 and BL$\,$Lac.
This range encompasses the displacements quoted by
Marscher et al. (1993), but shows that
the peculiar velocity of the foreground clouds has a significant
effect on the displacement, and hence, on our ability to 
``translate'' time-lapses into transverse displacements. 

The new 2009 observations reported here enable us to study variability
in a time lapse of almost two-decades.
From 1990 to 2009, the maximum estimated displacement is $\la 100\,$AU 
and as small as its parallax motion ($\la 1\,$AU; i.e., the case when 
the peculiar velocity of the cloud results in no proper motion)
for NRAO$\,$150, 3C111, and BL$\,$Lac. 
An estimate of the transverse displacement for B0212+735 is uncertain 
because there is no distance measurement to the foreground 
cloud reported in the literature, and kinematic distances (e.g., Watson et al. 2003) or 
statistical distances (e.g., Magnani et al. 1985) are unreliable
given the Galactic latitude and the small V$_{LSR}$ velocity of 
the absorption. However, a similar displacement range is expected 
for the foreground cloud in B0212+735 because the magnitude of 
its $V_{LSR}$ is within that of 3C111 and NRAO$\,$150 (Table~3), and the 
magnitude of its galactic latitude ($|b| = 12^o$) is similar to 
that of BL$\,$Lac ($|b| = 10.4^o$) and 3C111 ($|b| = 8.8^o$).

\subsection{Variability}

Figures~3 and 5 bottom right panels show the difference 
between the 2009 spectra and the previous observations for all sources
(i.e, ``red'' minus ``black'' spectra of the corresponding left 
panel), which will be called {\it residual spectra} hereafter. As mentioned 
above, clear variability is seen toward NRAO$\,150$ and 3C111 Core, however the
variability behavior observed toward both sources is quite different. 

In the case of 3C111 Core, we found no long-term variability
trend. The 2009 spectrum is the same within the noise
to the September 1991 and December 1990 spectra, and only one channel 
from the 1994 spectrum differs by $\sim$3.5$\sigma$ with respect 
to the 2009 spectrum. Moore \& Marscher (1995) reported variability of 
the H$_2$CO line from the July 1991
and December 1992 epochs with respect to their May 1994 spectrum. 
We also find variability above $3\sigma$ 
when comparing our 2009 spectrum with the July 1991 and December 1992
data (Figure~5). 
Comparing the March 2009 spectrum to the December 1992 spectrum 
(D92$-$M09 residual spectrum in Figure~5),
we see a line that resembles an inverse-P Cygni profile, which 
may be caused by a small velocity difference ($\sim 0.04$\kms; see Table~2)
between the two different sight-lines sampled on March 2009 and 
December 1992. 
Evidence for such velocity shift was pointed out by Moore \& Marscher (1995)
when comparing their 1992 and 1994 spectra.
The velocity difference is less than the FWHM of 
the line, thus, velocity gradients within the cloud could 
be responsible for the observed variability.
We note that the M94-M09 residual spectrum shows a similar profile than the D92-M09 
residual spectrum, albeit with significantly less signal to noise.
We do not detect evidence for such residual line profile in the
other epochs, nor toward the other sources, and at this point it is 
unclear why only the M92-M09 residual spectrum shows such a pronounced profile. 
Follow up observations of 3C111 are needed to detect more occurrences
of velocity shifts and detailed analysis of synthetic 
spectra from computer simulations of turbulent clouds is needed 
to estimate the expected occurrence rate of such events.

Figure~5 also shows that the peak of the 
July 1991 spectrum is less deep than the March 2009 spectrum by
$\Delta e^{-\tau} \sim 0.025$; Moore \& Marscher (1995) also report less
absorption in the July 1991 spectrum with respect to the May 1994 
spectrum by $\Delta e^{-\tau} \sim 0.02$ (see their Figure~5), however
the line in our residual spectrum is narrower than that from 
Moore \& Marscher (1995). The 3C111 residual spectra do not show evidence 
of hyperfine structure due to low signal-to-noise, which as mentioned 
in the introduction, would strengthen the case for variability.  

The difference between the July 1991 and March 2009 3C111 spectra could be 
explained by a change in the line-of-sight overall characteristics of the 
foreground cloud. 
Recalling that the main absorption feature in 3C111 Core is consistent with 
$n(H_2) = 10^4\,$cm$^{-3}$, $T_K = 30\,$K, $5\times10^{-9}$ H$_2$CO
abundance, $\Delta L = 10^4\,$AU, 
the difference between the July 1991 and March 2009 could be caused by
an $\sim$8\% variability in overall H$_2$CO abundance without 
changing any other parameter. Smaller abundance gradients would be needed
to explain the variability if the depth ($\Delta L$) of the H$_2$CO 
absorption region would also have changed. 
Significant H$_2$CO abundance gradients 
in molecular clouds are know to occur, e.g., Young et al. (2004) found
clear evidence of H$_2$CO depletion in the interior of pre-protostellar 
cores. 

In the case of NRAO$\,$150 (Figure~3), our new 2009 spectrum confirms the
variability trend reported by Moore \& Marscher (1995), i.e., the main 
absorption feature has been monotonically increasing since 1990. The case of
NRAO$\,$150 shows the advantages of using H$_2$CO absorption as a probe for
variability against point sources, in particular, the residual spectra show
detection of the $F=1-0$ hyperfine component and no variability of 
the $-17.3$\kms~line, 
which, as mentioned above, demonstrate that the residual lines are true 
variability and not artifacts. Figure~6 shows the variability as a function of time. 
The new data point shows a change in the slope of the variability, suggesting that 
the 1990's observations sampled a region with a greater column density/abundance 
gradient than the region sampled by the 2009 data. 

Assuming $n(H_2) = 10^4\,$cm$^{-3}$, $T_K = 30\,$K, $5\times10^{-9}$ H$_2$CO abundance, 
$\Delta L = 4 \times 10^3\,$AU and $N(H_2CO) = 3\times 10^{12}\,$cm$^{-2}$, 
the variability observed between 1990 and 2009 can be explained by an 
increase of $1.3\times 10^3\,$AU in the effective thickness of the molecular
cloud in the line of sight of the quasar ($\Delta L$) without any change in 
molecular density, temperature or abundance. As mentioned above, the transverse
displacement of the line of sight across the cloud between 1990 and 2009 is
$\la 100\,$AU. Assuming a transverse displacement of 50$\,$AU (half the 
maximum expected displacement), and 
that the cloud has a cylindrical filamentary structure (circular cross section) 
with the sight line to NRAO$\,$150 crossing the filament perpendicular to its length
(Figure~7), we can calculate the radius of curvature of the circular cross
section. We obtain a filament radius of $R \sim 2 \times 10^4\,$AU, 
and find that the line-of-sight would be crossing the filament very close to its 
edge (see Figure~7). We stress that the value of the radius is representative 
of the physical scales and not a precise measurement because of the large
uncertainties involved in $\Delta L$ and transverse displacement, and 
the filament is unlikely to have a perfectly circular cross-section.
Nevertheless, the molecular observations by Pety et al. (2008) indeed show that the line of sight
toward NRAO$\,$150 is crossing a filamentary structure, which 
has projected width of $\sim 40$\arcsec~(see their Figure~4, $-12 < V < -9.5\,$\kms~
range, 5.8\arcsec~resolution), which at a distance of 
700$\,$pc would correspond to a
radius of $\sim 1.4\times10^4\,$AU. This value is similar to the
radius of $\sim 2 \times 10^4\,$AU calculated above. 
Thus, without changing any other parameter of the overall physical conditions
of the gas (e.g., temperature, abundance, density) the variability observed
toward NRAO$\,$150 could be explained by a simple geometrical effect
caused by the line-of-sight crossing the edge of a molecular filament.

In order to explain the variability in H$_2$CO absorption profiles, 
Marscher et al. (1993) and Moore \& Marscher (1995) argue for the presence
of high density ($10^6\,$cm$^{-3}$) molecular clumps at scales of $\sim 10\,$AU.
In contrast, here we have found that minor chemical (abundance) gradients and/or
geometrical effects can account for the observed variability without the need
for high density clumps. Also, high density clumps are not necessarily
present in the case of foreground clouds that show no variability, such as 
toward B0212+735 or the $-17.2$\kms~component toward NRAO$\,$150. 
Moreover, we note that at such high densities 
($10^6\,$cm$^{-3}$) the H$_2$CO would be detected in emission instead of 
absorption against the CMB, which has never been observed in diffuse 
clouds.\footnote{Thermal emission of the 6$\,$cm H$_2$CO line has only been seen toward 
the Orion BN/KL region (Zuckerman et al. 1975, see also Araya et al. 2006) 
and non-thermal emission has been detected only toward young massive 
stellar objects (e.g., Araya et al. 2008).}
Hence, we concur with the conclusions Liszt \& Lucas (2000), that presented 
independent arguments against explaining absorption variability as caused by 
AU-sized high density inclusions in interstellar clouds. In particular, 
Liszt \& Lucas (2000) found that the high-density clump/inclusion assumption
is inconsistent with the weak HCO$^+$ emission in diffuse clouds.
As discussed in their paper, in order to explain the weak HCO$^+$ emission, a 
low ratio of total (high-density) clump to cloud area would be needed, but
this would be inconsistent with the frequent optical depth variations observed
in HCO$^+$ profiles.

\section{Summary}

We used the Very Large Array (VLA) in 2009 to obtain a new epoch in the monitoring of
6$\,$cm H$_2$CO lines from Galactic molecular
clouds detected in absorption toward extragalactic radio sources. We observed Galactic
absorption from diffuse/translucent molecular clouds toward B0212+735, 
NRAO$\,$150, 3C111, and BL$\,$Lac. The goal of the project was to check for
variability in the absorption lines due to transverse displacement of 
foreground clouds with respect to our line of sight to the extragalactic continuum sources; 
a displacement that is caused by the combined motion of the Solar System and the 
foreground clouds around the Galaxy. 

We detected 6$\,$cm H$_2$CO absorption toward
all sources. To check for the effect of different data reduction (calibration and 
imaging) procedures in observed spectra, we reduced the archived
observations reported by Marscher et al. (1993) and Moore \& Marscher (1995).
Our new spectra, together with the observations reduced from the archive, 
demonstrate the superb stability of the VLA for long term variability studies
of absorption lines. The data reported here, in particular the spectra obtained
toward NRAO$\,$150, show the advantages of using 6$\,$cm 
H$_2$CO absorption for this type of variability studies given the detection
of multiple absorption lines in the same line of sight and the presence of the
$F = 1-0$ hyperfine component that can be used to rule out spurious variability
in the main line. 

We did not detect significant variability toward 
B0212+735 and BL$\,$Lac.
Variability in the archive observations of 3C111 is seen which
can be explained by relatively minor chemical (abundance), 
velocity and/or cloud-thickness 
gradients in the foreground cloud. 
In the case of NRAO$\,$150, we confirm the monotonic increase in H$_2$CO 
absorption reported by Moore \& Marscher (1995). 
We found that the monotonic variability can be explained
by a filamentary (cylindrically)-shaped cloud crossing the line of sight to the
quasar, i.e., the variability could be caused by simple
geometrical effects without the need of high density molecular substructure 
(clumps or inclusions) in the medium, in agreement with Liszt \& Lucas (2000).
If the line of sight toward NRAO$\,$150 is probing the edge of a foreground
molecular filament as suggested by our analysis and as seen 
in CO maps of the region
(Pety et al. 2008), then, continuing monitoring of the H$_2$CO 
absorption toward this
radio source is important to further study the structure 
of the filament and to
explore other effects such as H$_2$CO depletion at higher extinctions across 
the cloud.

\acknowledgments

The National Radio Astronomy Observatory
is a facility of the National Science Foundation
operated under cooperative agreement by Associated Universities,
Inc. We acknowledge the anonymous referee for a careful review of the
manuscript and insightful suggestions. This work has made use of the 
computational facilities donated by Frank Rodeffer to the Astrophysics 
Research Laboratory of Western Illinois University. E.D.A. acknowledges 
support from the Provost Travel Award and the WIU Physics 
Department to present some of these results at an AAS meeting. 
This research has made use of NASA's Astrophysics 
Data System, and the NASA/IPAC Extragalactic Database (NED) which is operated 
by the Jet Propulsion Laboratory, California Institute of Technology, 
under contract with the National Aeronautics and Space Administration.

\clearpage

\begin{deluxetable}{lcccccc}
\tabletypesize{\scriptsize}
\tablewidth{0pt}
\tablecaption{Observation Summary}
\tablehead{
\colhead{Line of Sight} & \colhead{Run$^+$} & \colhead{R.A. (J2000)} &
\colhead{Decl. (J2000)} & \colhead{BW} & \colhead{$\Delta v$} &
\colhead{Synthesized Beam} \\
\colhead{} & \colhead{} & \colhead{(h m s)} & 
\colhead{(\arcdeg~\arcmin~\arcsec)} & 
\colhead{(\kms)} & \colhead{(\kms)} & 
\colhead{($\theta_{max} \times \theta_{min}$; P.A.)} }
\startdata
B0212+735   & M09 & 02 17 30.813 & +73 49 32.62 & 96.8 & 0.19  & 2.7\arcsec  $\times$ 1.4\arcsec,    83\arcdeg \\
            & M94 & 02 17 30.819 & +73 49 32.67 & 12.0 & 0.095 & 0.69\arcsec $\times$ 0.44\arcsec,   41\arcdeg \\
            & D92 & 02 17 30.819 & +73 49 32.67 & 12.0 & 0.095 & 0.60\arcsec $\times$ 0.40\arcsec,   14\arcdeg \\
NRAO$\,$150 & M09 & 03 59 29.747 & +50 57 50.16 & 96.8 & 0.19  & 1.3\arcsec  $\times$ 1.1\arcsec, $-$16\arcdeg \\
            & M94 & 03 59 29.746 & +50 57 50.22 & 12.0 & 0.095 & 0.56\arcsec $\times$ 0.47\arcsec,   56\arcdeg \\ 
            & D92 & 03 59 29.746 & +50 57 50.22 & 12.0 & 0.095 & 0.48\arcsec $\times$ 0.40\arcsec,    2\arcdeg \\ 
            & J91 & 03 59 29.742 & +50 57 50.13 & 20.0 & 0.095 & 0.52\arcsec $\times$ 0.41\arcsec,   45\arcdeg \\ 
            & D90 & 03 59 29.746 & +50 57 50.22 & 48.3 & 0.19  & 5.5\arcsec  $\times$ 4.5\arcsec,     3\arcdeg \\
3C111       & M09$^*$ & 04 18 21.277 & +38 01 35.80 & 96.8 & 0.19  & 2.3\arcsec  $\times$ 1.4\arcsec, $-$87\arcdeg \\
            & M94 & 04 18 21.299 & +38 01 35.77 & 12.0 & 0.095 & 0.50\arcsec $\times$ 0.46\arcsec,   24\arcdeg \\
            & D92 & 04 18 21.299 & +38 01 35.77 & 12.0 & 0.095 & 0.45\arcsec $\times$ 0.40\arcsec, $-$6\arcdeg \\
            & S91 & 04 18 21.298 & +38 01 35.72 & 12.0 & 0.095 & 0.52\arcsec $\times$ 0.46\arcsec,   70\arcdeg \\ 
            & J91 & 04 18 21.298 & +38 01 35.72 & 12.0 & 0.095 & 0.48\arcsec $\times$ 0.44\arcsec, $-$5\arcdeg \\ 
            & D90 & 04 18 22.915 & +38 01 43.27 & 48.3 & 0.19  & 5.5\arcsec  $\times$ 4.7\arcsec,  $-$9\arcdeg \\
BL$\,$Lac   & M09 & 22 02 43.291 & +42 16 39.98 & 96.8 & 0.19  & 2.4\arcsec  $\times$ 1.7\arcsec,    63\arcdeg \\
            & M94 & 22 02 43.279 & +42 16 40.05 & 12.0 & 0.095 & 0.53\arcsec $\times$ 0.43\arcsec, $-$10\arcdeg \\
            & D92 & 22 02 43.279 & +42 16 40.05 & 12.0 & 0.095 & 0.49\arcsec $\times$ 0.42\arcsec, $-$55\arcdeg \\
            & S91 & 22 02 43.288 & +42 16 39.89 & 12.0 & 0.095 & 0.49\arcsec $\times$ 0.42\arcsec,    7\arcdeg  \\
            & J91 & 22 02 43.288 & +42 16 39.89 & 12.0 & 0.095 & 0.47\arcsec $\times$ 0.40\arcsec,  $-$6\arcdeg \\
            & D90 & 22 02 43.279 & +42 16 40.05 & 48.3 & 0.19  & 5.2\arcsec  $\times$ 4.8\arcsec,     34\arcdeg \\
\enddata
\tablecomments{\scriptsize The right ascension and declination
of the phase tracking center are given (coordinates were precessed to 
J2000 equinox), as well as the bandwidth
and initial channel width (all spectra were smoothed to a channel 
width of 0.19\kms~for the analysis). Note that in the case of the
2009 observations, only $\sim 35\%$ of the bandwidth could 
be used due to aliasing.
($^+$) The different runs correspond to 02 March 2009 (M09), 01 May 1994 (M94), 
27 December 1992 (D92), 13 July 1991 (J91), 10 December 1990 (D90). The VLA project 
number of the 2009 observations is AA327, all other runs correspond to 
observations published by Moore \& Marscher (1995; VLA archive project numbers
AM309, AM331, AM376, AM440).
($^*$) On the M09 run, we observed 3C111 also in continuum mode, 
which resulted in a synthesized beam = 3.2\arcsec  $\times$ 1.4\arcsec, P.A. = $-$70\arcdeg.}
\end{deluxetable}

\clearpage

\begin{deluxetable}{lccccc}
\tabletypesize{\scriptsize}
\tablewidth{0pt}
\tablecaption{Line Parameters}
\tablehead{
\colhead{Line of Sight} & \colhead{Run} & 
\colhead{$e^{-\tau}$} &  \colhead{rms$_{e^{-\tau}}$} & 
\colhead{$V_{LSR}$} & \colhead{$FWHM$} \\
\colhead{} & \colhead{} & 
\colhead{} & \colhead{} & 
\colhead{(\kms)} & \colhead{(\kms)}}
\startdata
B0212+735    & M09     &  0.928   & 0.002 	&     3.39  (0.01)  & 1.06 (0.03) \\ 
             & M94     &  0.932   & 0.003 	&     3.42  (0.01)  & 1.11 (0.04) \\ 
             & D92     &  0.937   & 0.004 	&     3.37  (0.04)  & 1.08 (0.09) \\ 
NRAO$\,$150  & M09$^1$ &  0.810   & 0.002 	& $-$10.57  (0.02)  & 0.70 (0.04) \\ 
             & M94     &  0.840   & 0.003 	& $-$10.54  (0.01)  & 0.69 (0.02) \\ 
             & D92     &  0.852   & 0.003 	& $-$10.53  (0.01)  & 0.73 (0.02) \\ 
             & J91     &  0.860   & 0.003 	& $-$10.52  (0.02)  & 0.69 (0.04) \\ 
             & D90     &  0.870   & 0.005 	& $-$10.53  (0.02)  & 0.76 (0.04) \\ 
             & M09     &  0.942   & 0.002 	& $-$17.30  (0.02)  & 0.76 (0.05) \\ 
             & J91     &  0.934   & 0.003 	& $-$17.30  (0.02)  & 0.69 (0.05) \\ 
             & D90     &  0.943   & 0.005 	& $-$17.3   (0.1)   & 0.8  (0.3)  \\ 
             & M09     &  0.984   & 0.002 	& $-$8.6    (0.1)   & 1.1  (0.2)  \\ 
             & M94     &  0.984   & 0.003 	& $-$8.6    (0.1)   & 1.0  (0.2)  \\ 
             & D92     &  0.986   & 0.003 	& $-$8.4    (0.2)   & 0.7  (0.2)  \\
             & J91     &  0.987   & 0.003 	& $-$8.4    (0.1)   & 0.7  (0.2)  \\
             & D90     &  0.989   & 0.005 	& $-$8.4    (0.1)   & 0.4  (0.3)  \\
3C111-CORE   & M09     &  0.677   & 0.002 	&  $-$1.01  (0.01)  & 0.90 (0.01) \\ 
             & M94     &  0.677   & 0.002 	&  $-$1.0   (0.2)   & 1.0  (0.2)  \\ 
             & D92     &  0.685   & 0.003 	&  $-$0.97  (0.05)  & 0.95 (0.2)  \\ 
             & S91     &  0.684   & 0.007 	&  $-$1.0   (0.2)   & 1.0  (0.2)  \\ 
             & J91     &  0.695   & 0.005 	&  $-$1.01  (0.05)  & 0.94 (0.03) \\ 
             & D90     &  0.675   & 0.005 	&  $-$1.0   (0.1)   & 0.90 (0.06) \\ 
             & M09     &  0.960   & 0.002 	& $-$2.29   (0.07)  & 1.8  (0.1)  \\ 
             & M94     &  0.958   & 0.002       & $-$2.6    (0.2)   & 1.3  (0.2)  \\ 
             & D92     &  0.960   & 0.003 	& $-$2.4    (0.1)   & 1.6  (0.1)  \\ 
             & S91     &  0.970   & 0.007 	& $-$2.7    (0.2)   & 1.3  (0.2)  \\ 
             & J91     &  0.960   & 0.005 	& $-$2.4    (0.1)   & 1.6  (0.3)  \\ 
             & D90     &  0.956   & 0.005 	& $-$2.4    (0.1)   & 1.6  (0.2)  \\ 
3C111-NE     & M09     &  0.70    & 0.01  	&  $-$1.15  (0.02)  & 0.89 (0.04) \\ 
             & M94     &  0.71    & 0.03 	&  $-$1.1   (0.2)   & 0.8  (0.2)  \\ 
             & D92     &  0.68    & 0.05 	&  $-$1.19  (0.04)  & 0.6  (0.2)  \\ 
             & S91     &  0.56    & 0.09 	&  $-$1.13  (0.07)  & 1.0  (0.2)  \\ 
             & J91     &  0.60    & 0.07 	&  $-$1.10  (0.05)  & 0.7  (0.1)  \\ 
             & D90     &  0.69    & 0.01 	&  $-$1.1   (0.2)   & 0.9  (0.2)  \\ 
             & M09     &  0.93    & 0.01  	&  $-$2.46  (0.05)  & 1.3  (0.2)  \\ 
             & M94     &  0.94    & 0.03 	&  $-$2.3   (0.2)   & 1.5  (0.2)  \\ 
             & D90     &  0.91    & 0.01 	&  $-$2.5   (0.2)   & 1.2  (0.2)  \\ 
3C111-SW     & M09     &  0.68    & 0.05  	&  $-$1.67  (0.07)  & 1.4  (0.2)  \\ 
             & M94     & \nodata  & 0.4   	&    \nodata        & \nodata     \\
             & D92     & \nodata  & 0.3   	&    \nodata        & \nodata     \\
             & S91     & \nodata  & 0.6   	&    \nodata        & \nodata     \\
             & J91     & \nodata  & 0.4   	&    \nodata        & \nodata     \\
             & D90     &  0.83    & 0.06  	&  $-$1.7   (0.1)   & 2.0  (0.3)  \\ 
BL$\,$Lac$^2$& M09     &  0.957   & 0.002 	&  $-$1.57  (0.03)  & 0.95 (0.07) \\ 
             & M94     &  0.958   & 0.002       &  $-$1.62  (0.03)  & 0.93 (0.05) \\ 
             & D92     &  0.961   & 0.002       &  $-$1.66  (0.04)  & 0.88 (0.07) \\ 
             & S91     &  0.958   & 0.002 	&  $-$1.56  (0.02)  & 0.88 (0.04) \\ 
             & J91     &  0.959   & 0.002 	&  $-$1.55  (0.02)  & 1.01 (0.05) \\ 
             & D90     &  0.957   & 0.004 	&  $-$1.62  (0.07)  & 1.0  (0.1)  \\
             & M09     &  0.984   & 0.002 	&  $-$0.41  (0.06)  & 1.11 (0.14) \\ 
             & M94     &  0.984   & 0.002       &  $-$0.46  (0.07)  & 1.06 (0.18) \\ 
             & D92     &  0.984   & 0.002 	&  $-$0.5   (0.1)   & 1.3  (0.3)  \\ 
             & S91     &  0.983   & 0.002 	&  $-$0.52  (0.04)  & 0.7  (0.1)  \\ 
             & J91     &  0.986   & 0.002 	&  $-$0.37  (0.05)  & 0.8  (0.1)  \\ 
             & D90     &  0.987   & 0.004 	&  $-$0.5   (0.2)   & 1.0  (0.4)  \\
\enddata
\tablecomments{
\scriptsize 
Unless indicated otherwise, the line parameters were obtaining by fitting 
the main group of five hyperfine components with a single Gaussian profile; the $F=1-0$ 
hyperfine component was excluded from the fit. 
$^{(1)}$ Gaussian fit underestimates optical depth. The peak $e^{-\tau}$ and
channel velocity are 0.790 and $-$10.48\kms; see Table~3 for fit including hyperfine structure.  
$^{(2)}$ Complex absorption profile fitted with two Gaussians. The weaker Gaussian profile overlaps 
with the $F=1-0$ hyperfine component of the strong line.}
\end{deluxetable}

\clearpage

\begin{deluxetable}{lccc}
\tabletypesize{\scriptsize}
\tablewidth{0pt}
\tablecaption{Hyperfine Fit of 2009 Spectra}
\tablehead{
\colhead{Line of Sight} & 
\colhead{$\tau_\nu$ (F=2-2)} & 
\colhead{$V_{LSR}$} & \colhead{$FWHM$} \\ 
\colhead{} & \colhead{} & 
\colhead{(\kms)} & \colhead{(\kms)} 
}
\startdata
B0212+735     & 0.044 (0.001) &     3.54 (0.01)  & 0.82 (0.03)  \\
NRAO$\,$150   & 0.17  (0.03)  & $-$10.47 (0.03)  & 0.40 (0.07)  \\
              & 0.05  (0.02)  & $-$17.2  (0.1)   & 0.4  (0.3)   \\ 
3C111-CORE$^1$& 0.243 (0.004) &  $-$0.87 (0.01)  & 0.65 (0.01)  \\
3C111-NE      & 0.24  (0.01)  &  $-$1.01 (0.01)  & 0.56 (0.02)  \\
3C111-SW      & 0.20  (0.02)  &  $-$1.51 (0.05)  &  1.1 (0.1)   \\
BL$\,$Lac$^2$ & 0.023 (0.009) & $-$1.54 (0.10)   & 0.57 (0.25)  \\
              & 0.010 (0.005) & $-$0.80 (0.23)   & 0.96 (0.55)  \\
\enddata
\tablecomments{
\scriptsize Line parameters obtained by fitting the six hyperfine 
components of the H$_2$CO line. Resulting fits are shown in 
Figures 1 to 5 (blue dashed curves).  
LTE line intensities 
were assumed (Tucker et al. 1970). The optical depth
corresponds to the $F=2-2$ transition. 
1$\sigma$ errors from the fit are shown in parenthesis. 
$^{(1)}$ A second velocity component overlaps with the main absorption line. 
The line parameters of the overlapping line are 
$V_{LSR} = -2.29$\kms, $FWHM = 1.8$\kms, $\tau = 0.0356$.
$^{(2)}\,$A reliable hyperfine fitting of the line could not be 
obtained due to overlapping velocity components. We report the 
hyperfine line parameters assuming two velocity components. However,
a third velocity component is still needed to account for the 
absorption at 0.3\kms~(Figure~2, upper right panel).} 
\end{deluxetable}

\clearpage

\begin{figure}
\includegraphics{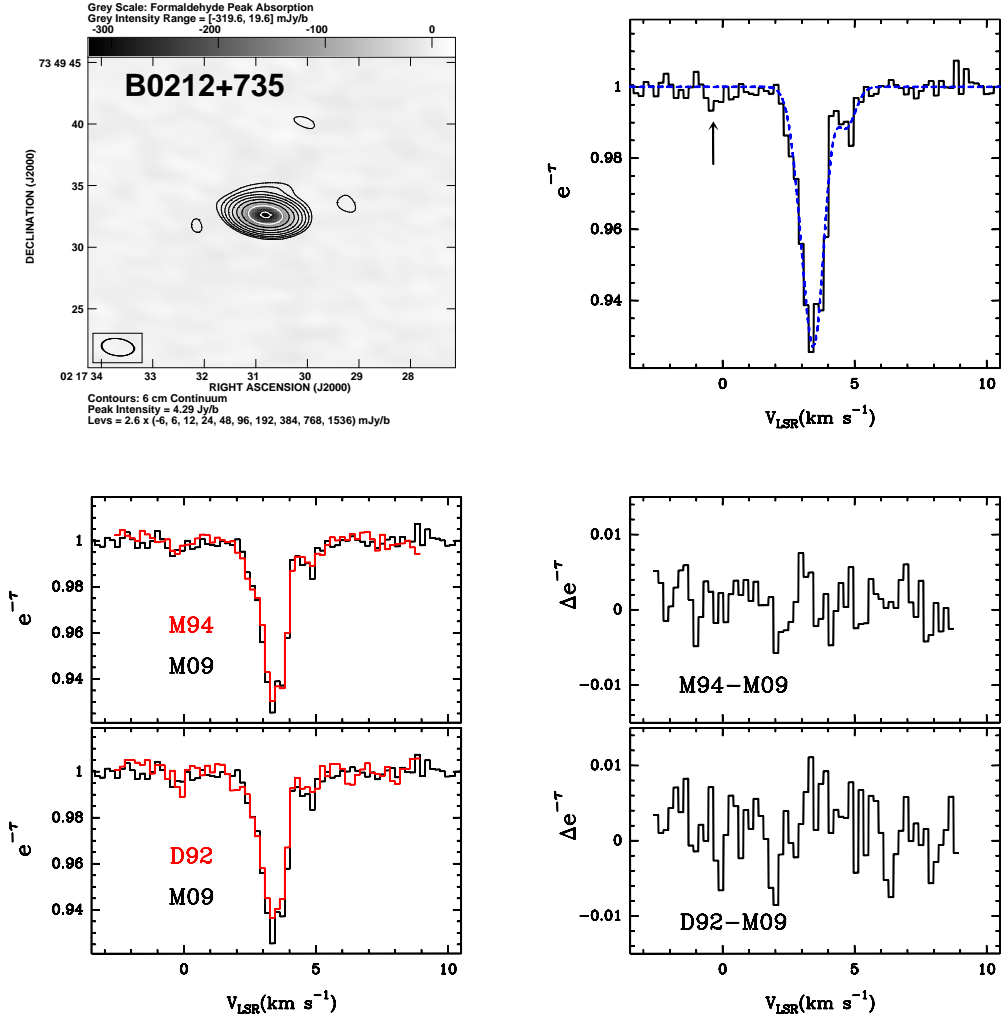} 
\vspace*{16cm}\caption{{\it Upper left panel:} Radio continuum image of the extragalactic
radio source B0212+735 is shown in contours; the gray scale shows the distribution of the 
peak H$_2$CO absorption toward the quasar (minimum value of the H$_2$CO spectrum in each pixel). 
{\it Upper right panel:} Spectrum obtained in the March 2009 run of this 
project. The blue dashed line is the fit to the
six hyperfine components of the transition (Table~3). The arrow highlights a weak absorption feature.
{\it Middle} and {\it bottom left panels} show the M09 spectra (black) and the
spectra from the other epochs (red); the difference between the red and black spectra
are presented in the {\it middle} and {\it bottom right panels}.}
\label{f1}
\end{figure}

\begin{figure}
\includegraphics{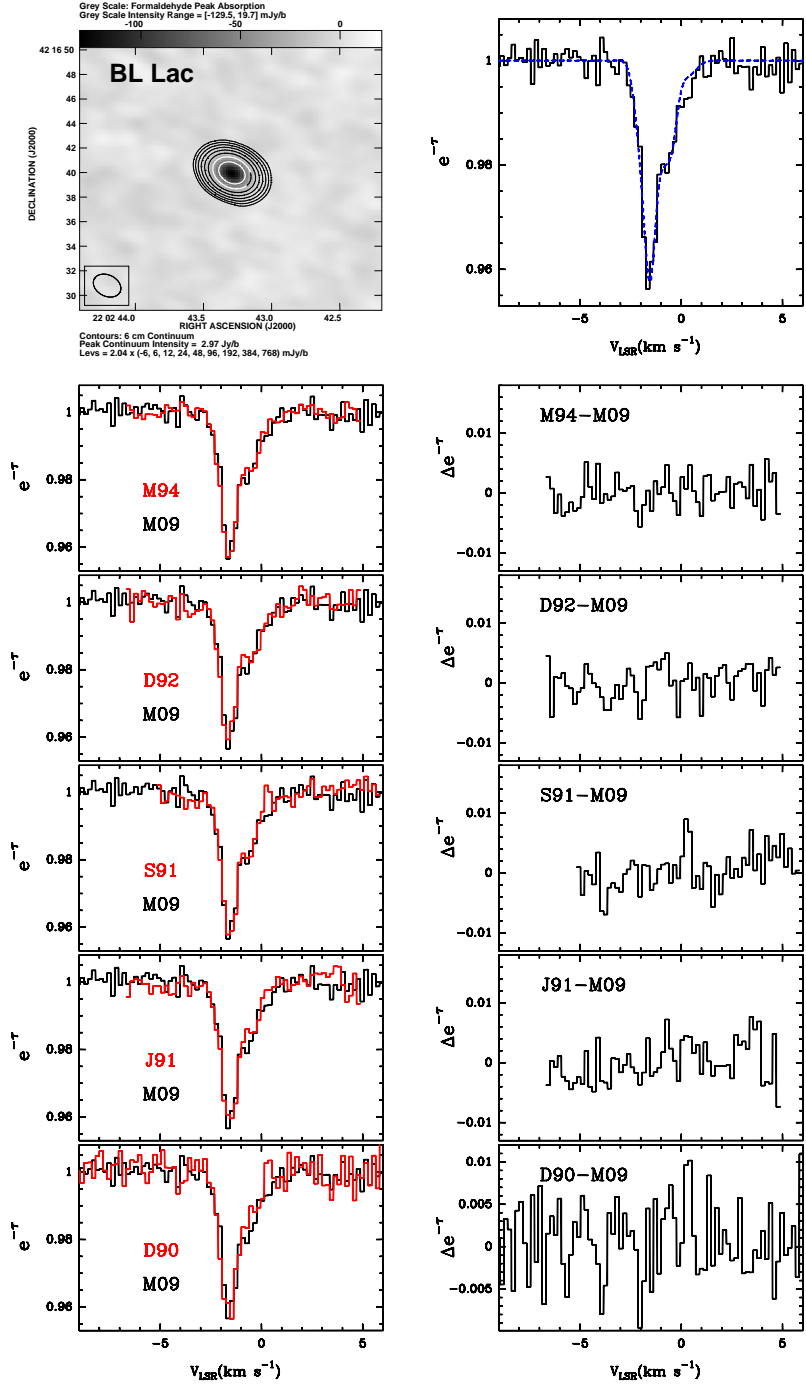} 
\vspace*{18cm}\caption{Same as Figure~1 for BL$\,$Lac. The blue dashed 
line shown in the upper right panel is the hyperfine line fit of 
two overlapping velocity components (see Table~3). A third velocity component
($e^{-\tau} \approx 0.993$) is needed to explain the absorption at $\sim$0.3\kms.}
\label{f1}
\end{figure}

\begin{figure}
\includegraphics{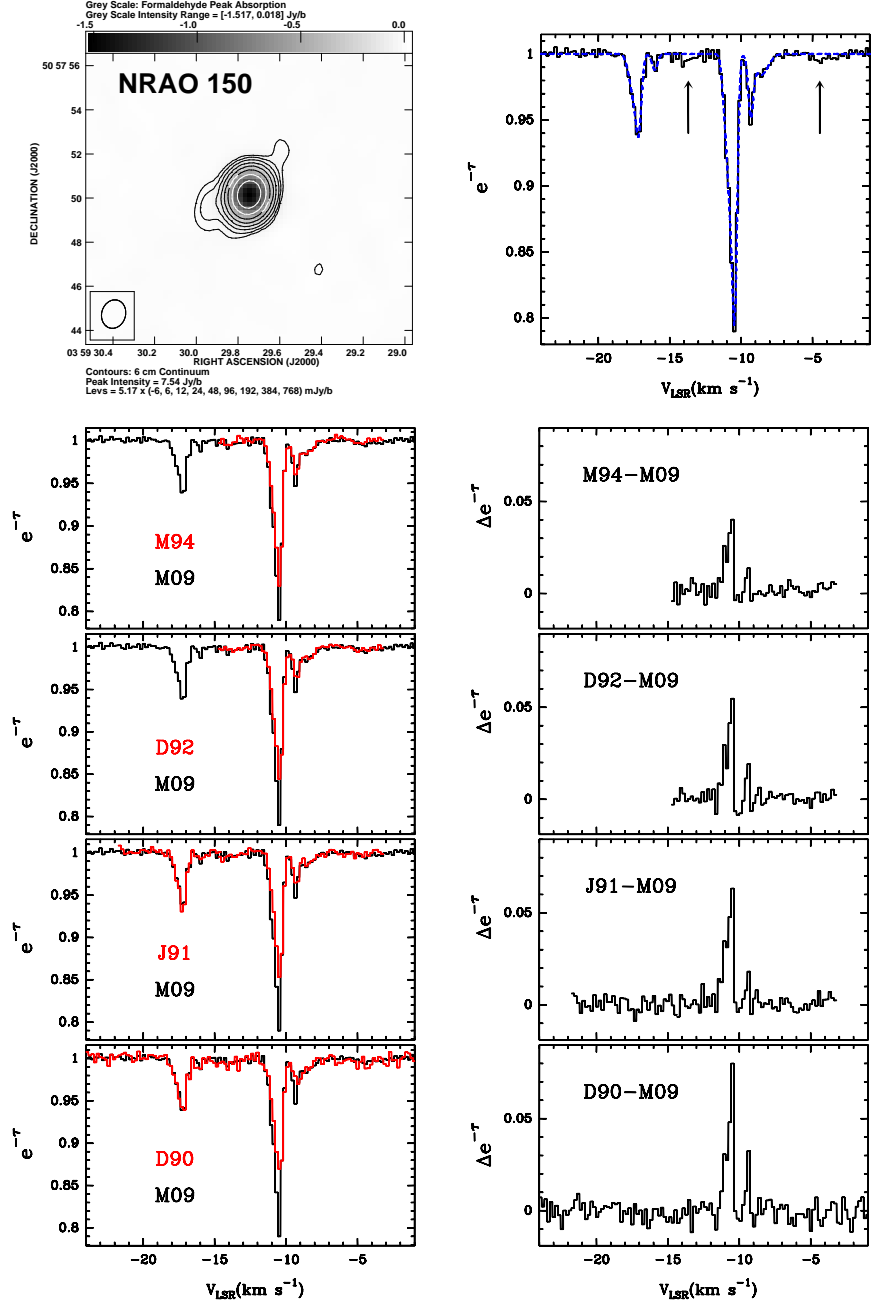} 
\vspace*{17cm}\caption{Same as Figure~1 for NRAO$\,$150.
The blue dashed line is the fit to the six hyperfine components of the transition (Table~3),
however the (broad and weak) line at $-$8.6\kms~is not part of the hyperfine structure of the main 
absorption line. The line parameters of the $-$8.6\kms~line are given in Table~2. The narrow 
line at $\sim -$9\kms~is the $F=1-0$ hyperfine component of the main line.
The arrows highlight weak absorption features.}
\label{f1}
\end{figure}

\begin{figure}
\includegraphics{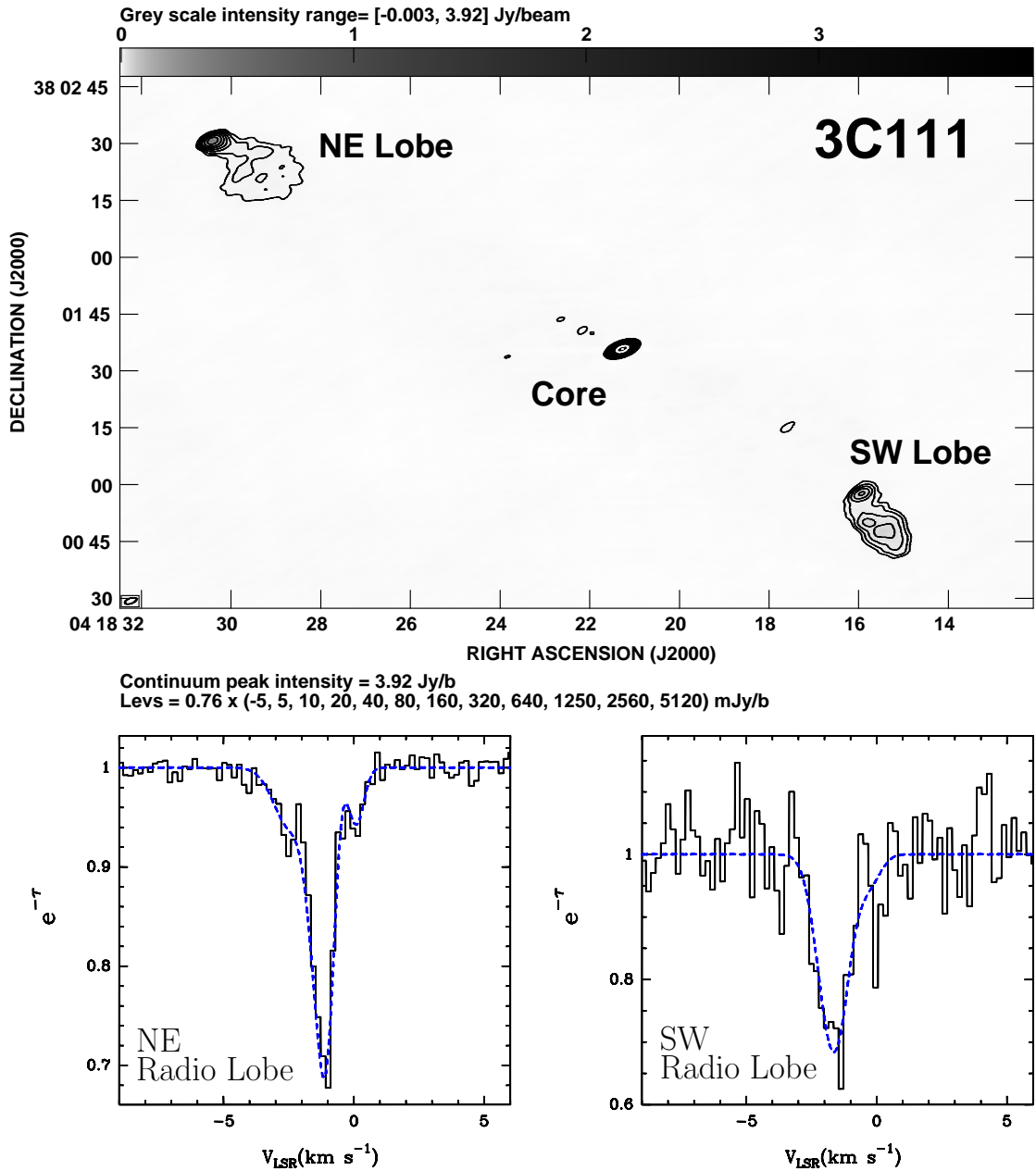} 
\vspace*{15cm}\caption{{\it Upper Panel:} Radio continuum image of 3C111. This image was obtained from the continuum set up observations and was used as initial model for self-calibration of the line data. {\it Lower Panels:} Spectra obtained in the March 2009 run toward the NE and SW radio lobes of 3C111. As in Figure~1, the blue dashed line is the fit to the six hyperfine components of the transition with the exception of the overlapping $-$2.46\kms~velocity component (NE 3C111 spectrum) whose parameters are reported in Table~2.}
\label{f_cont}
\end{figure}

\begin{figure}
\includegraphics{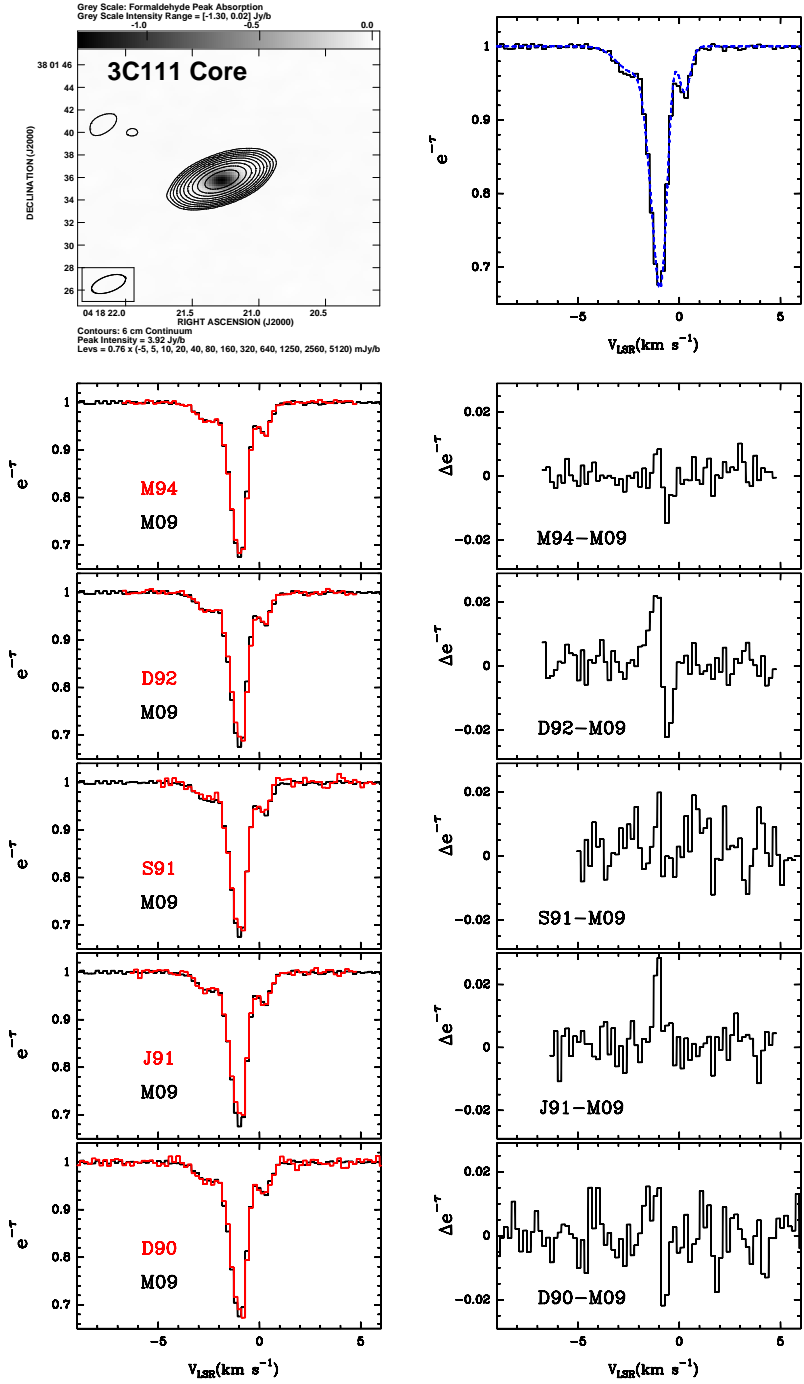} 
\vspace*{18cm}\caption{Same as Figure~1 for the 3C111 Core. Continuum is from Figure 4. A second velocity component at $\sim -$2.3\kms~overlaps with the main absorption line (see Table~2). Note the ``inverse-P Cygni'' profile of the D92-M09 residual spectrum which may be caused by a slight velocity difference between the D92 and M09 peaks (see Table~2 and Section 4.4).}
\label{f1}
\end{figure}

\begin{figure}
\includegraphics{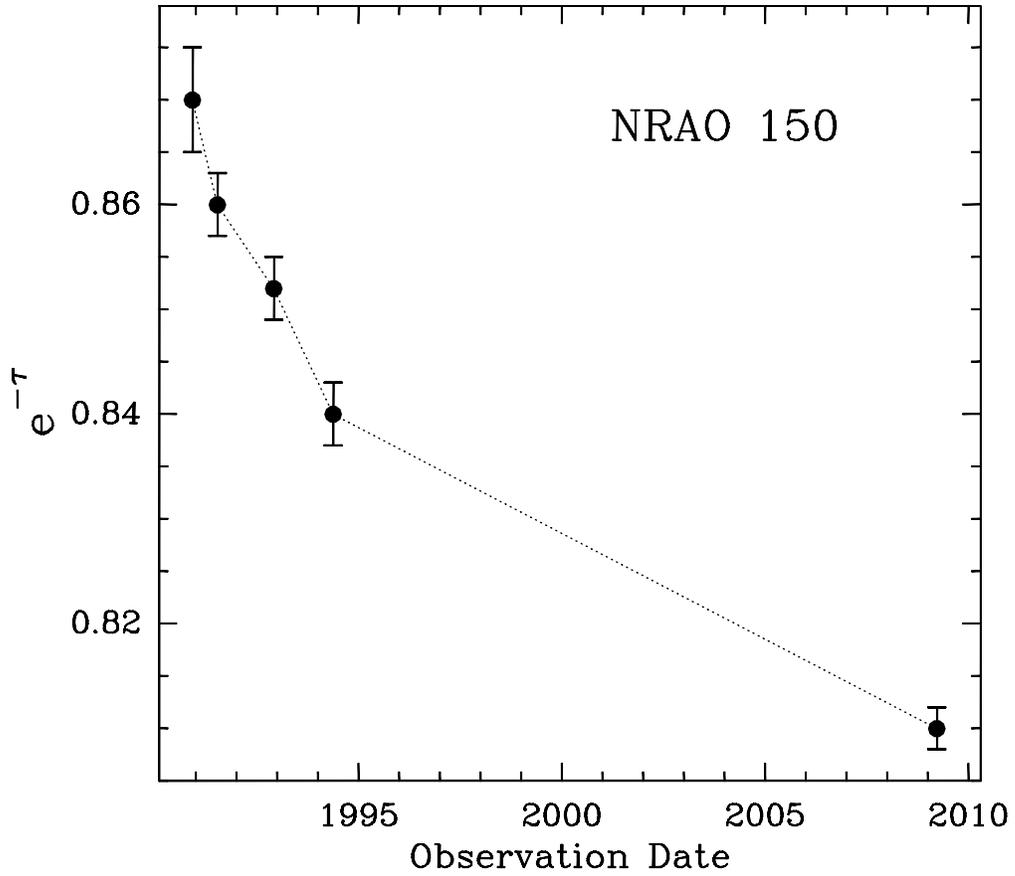}
\vspace*{17cm}\caption{Variability of the peak H$_2$CO absorption line in NRAO$\,$150 as 
a function of time. Our results confirm a monotonic increase in absorption, 
which suggests that the physical scale of the substructure sampled by the data is 
greater than the transverse displacement over the time period.}
\label{f1}
\end{figure}

\begin{figure}
\includegraphics{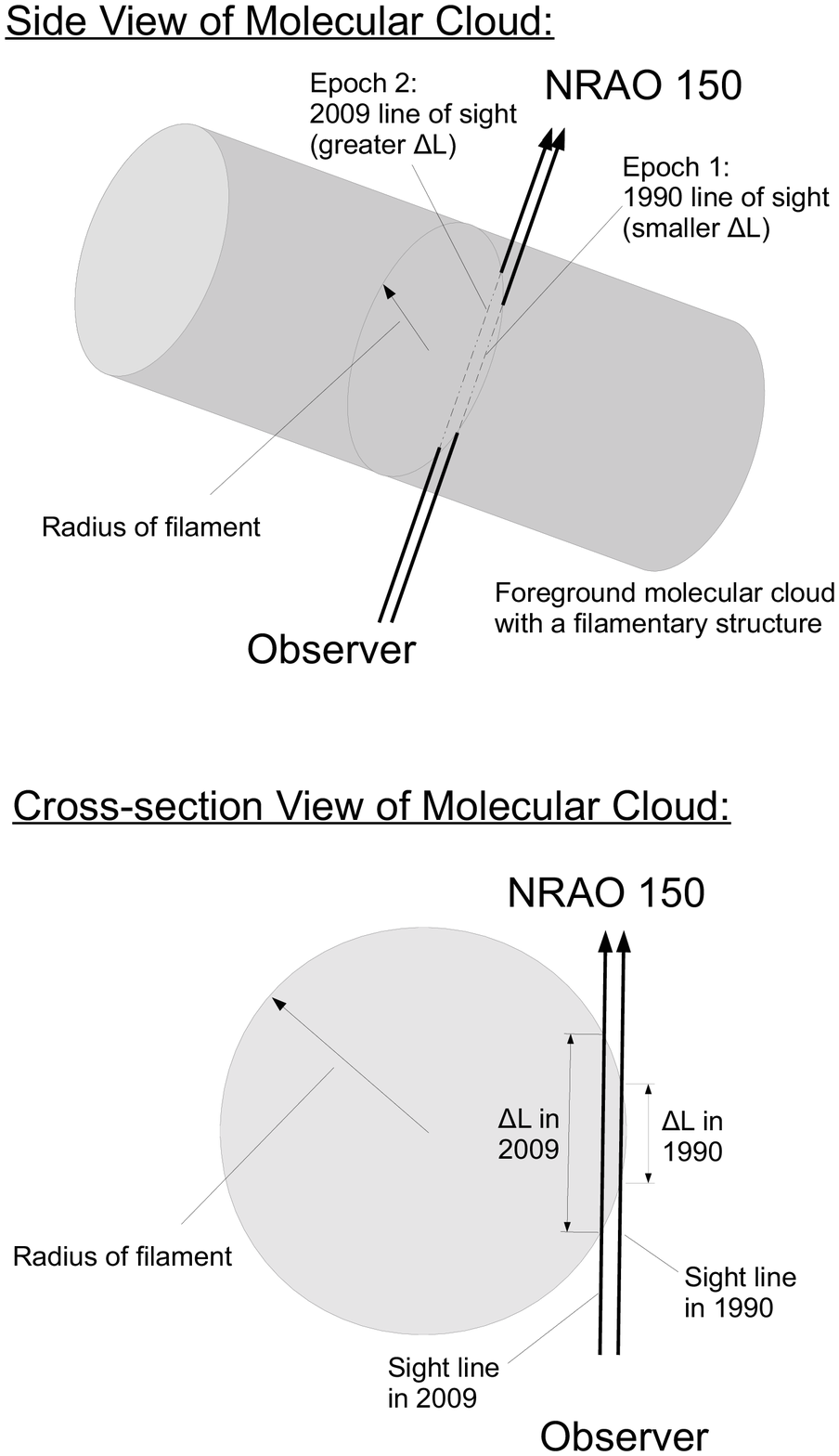}
\vspace*{17cm}\caption{Scheme of the foreground molecular cloud geometry used 
to explain the variability of the H$_2$CO absorption profile toward NRAO$\,$150. 
We assume that the transverse displacement of the line of sight toward the quasar 
is parallel to the cross section of the filament and moving inwards, i.e., 
the absorption as been increasing as a function of time, thus the line of 
sight path across the filament ($\Delta L$) has been increasing since 1990.}
\label{f1}
\end{figure}

\end{document}